\documentclass[
floatfix,
aps,
pra,
amsmath,
twocolumn,
superscriptaddress
]{revtex4-2}

\pdfoutput=1

\usepackage{xtab}
\usepackage{graphicx}
\usepackage{wrapfig}
\usepackage{amsmath}
\usepackage{amssymb}
\usepackage{bbm}
\usepackage{hyperref}
\usepackage{soul}
\usepackage{xcolor}
\usepackage{bbold}
\usepackage{multirow}
\usepackage{amsfonts}
\usepackage{booktabs}
\usepackage{siunitx}
\usepackage{booktabs}
\usepackage{amsmath}
\usepackage{array}
\usepackage[table,xcdraw]{xcolor}
\usepackage{float}
\usepackage{orcidlink}
\usepackage{svg}
\usepackage[table]{xcolor}

\newcommand{\ket}[1]{|{#1}\rangle}
\newcommand{\bra}[1]{\langle{#1}|}

\begin{document}

\title{CNOT gates in inductively coupled multi-fluxonium systems}

\author{Valeria Díaz Moreno}
\email{diazmoreno@wisc.edu}
\affiliation{Department of Physics, University of Wisconsin-Madison, Madison, WI 53706, USA}

\author{Nikola D. Dimitrov}
\affiliation{Department of Physics, University of Wisconsin-Madison, Madison, WI 53706, USA}
\affiliation{SEEQC, Inc., Elmsford, NY 10523, USA}

\author{Vladimir E. Manucharyan}
\affiliation{Institute of Physics, Ecole Polytechnique Federale de Lausanne, Lausanne, Switzerland}

\author{Maxim G. Vavilov}
\affiliation{Department of Physics, University of Wisconsin-Madison, Madison, WI 53706, USA}

\date{\today}

\begin{abstract}
High-fidelity two-qubit gates have been demonstrated in systems of two fluxonium qubits; however, the realization of scalable quantum processors requires maintaining low error rates in substantially larger architectures. In this work, we analyze a system of four inductively coupled fluxonium qubits to determine the impact of spectator qubits on the performance of a \textsc{cnot} gate. Our results show that spectator-induced errors are strongly suppressed when the transition frequencies of the spectator qubits are sufficiently detuned from those of the active qubits. We identify favorable frequency configurations for the four-qubit chain that yield \textsc{cnot} gate errors below $10^{-4}$
for gate times shorter than 100 ns. Leveraging the locality of the nearest-neighbor coupling, we extrapolate our findings to longer fluxonium chains, suggesting a viable path toward scalable, low-error quantum information processing.

\end{abstract}

\maketitle

\section{Introduction}

Superconducting qubits \cite{Kjaergaard_2020, PRXQuantum.2.030101} have emerged as one of the leading qubit platforms, exhibiting compatibility with standard integrated circuit fabrication processes \cite{PhysRevLett.111.080502}, enabling the construction of large-scale quantum processors \cite{2024}, and supporting the implementation of a universal set of quantum gates \cite{long2021universal}. 
Within the family of superconducting qubit architectures, fluxonium qubits \cite{Manucharyan_2009}
have become a promising platform due to their long coherence times and enhanced anharmonicity \cite{PhysRevX.9.041041,PhysRevLett.130.267001,PhysRevLett.129.010502}. These characteristics are advantageous for fast, low-leakage multiqubit gates \cite{zhao2025scalable}.
Recent studies have demonstrated high-fidelity single and two-qubit gates in fluxonium-based architectures, achieving error rates below $10^{-4}$ \cite{PhysRevApplied.18.034027, Nesterov_2022, PRXQuantum.5.040342}. Additional approaches—such as the exploration of alternative coupling mechanisms \cite{rosenfeld2024designinghighfidelitytwoqubitgates}, the use of microwave drives that address transitions outside the nominal computational subspace \cite{PRXQuantum.2.020345}, and the implementation of tunable couplers \cite{Moskalenko_Simakov_Abramov_Grigorev_Moskalev_Pishchimova_Smirnov_Zikiy_Rodionov_Besedin_2022} have also produced promising gate performance.

Recently, a two-qubit high-fidelity gate has been realized in a two-fluxonium system with direct inductive coupling \cite{Lin_2025,Lin2025_PRXQ}. In addition to enabling a high-fidelity entangling operation, this setup exhibited long-term stability and did not require recalibration over several days. The inductive coupling further resulted in a low static ZZ interaction, even for moderate coupling strengths between the fluxonium qubits. This naturally raises the question of whether such an architecture can be systematically scaled beyond two qubits.

In this work, we investigate the performance of a controlled-NOT (\textsc{cnot}) gate implemented between two adjacent qubits in a linear array of inductively coupled fluxonium devices. Within this architectural paradigm, the presence of additional non-participating ``spectator'' qubits is expected to adversely affect the achievable gate fidelity. We aim to quantitatively characterize and evaluate this spectator-induced reduction in gate fidelity in such multiqubit configurations.

As a reference, we first study \textsc{cnot} gate implementations in an isolated two-qubit system composed of two coupled fluxonium devices, considering three distinct frequency arrangements of the control and target qubits: low–medium (LM), medium–high (MH), and low–high (LH). For all three configurations, we obtain gate error probabilities well below \(10^{-5}\) for gate durations of 100 ns.
We subsequently extend the system to a four-qubit chain by introducing two spectator qubits adjacent to the control and target qubits. Our analysis focuses on two principal scenarios: (i) assessing gate performance in the four-qubit setting across a range of low–medium–high (LMH) frequency configurations, and (ii) performing frequency sweeps of the spectator qubits while keeping the gate qubits fixed in an LM configuration.

\begin{figure}[htbp]
  \centering
  \includegraphics[width=0.5\textwidth]{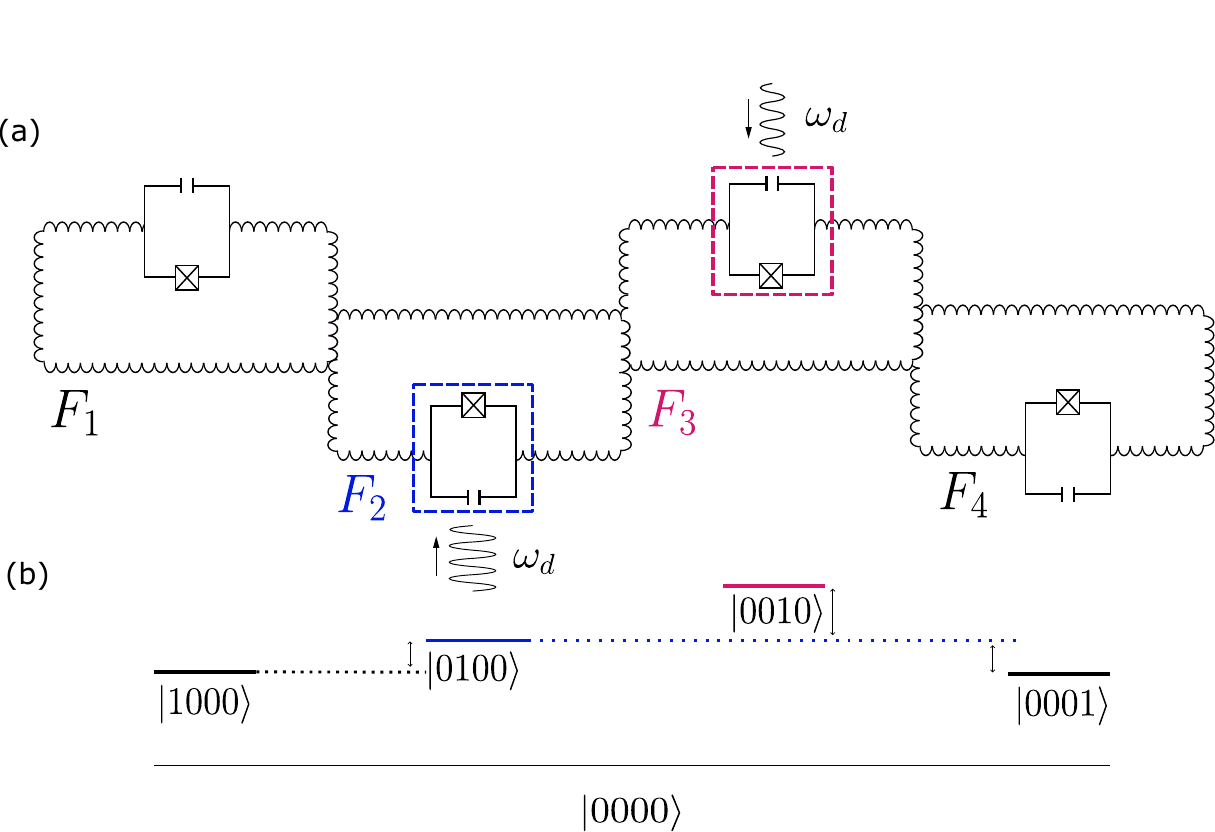}
  \caption{(a) Circuit diagram of four inductively coupled fluxonium qubits arranged linearly and labeled $F_1$, $F_2$, $F_3$, and $F_4$. A microwave drive at the target-qubit transition frequency $\omega_{d}$ is applied locally to the two central qubits, which serve as the control and target qubits during the gate operation. The two outer qubits function as spectators. (b) Energy-level diagram showing the computational subspace of the four-qubit chain. Detuning among qubits is indicated by the relative separation of the single-excitation energy levels.}
  \label{fig:fluxsyst}
\end{figure}

Using microwave pulse parameters optimized for the two-qubit gate in the extended four-qubit system, we observe a notable degradation in the gate performance. We subsequently re-optimize the microwave pulse parameters for the whole four-qubit system, which improves the \textsc{cnot} fidelity but does not fully recover the performance of the isolated two-qubit gate. Nevertheless, for the four-qubit system, we obtain gate errors below $10^{-4}$ for gate times under 100 ns, with select frequency configurations exhibiting significantly better performance than others.

Notably, we find that the gate performance in a four-qubit system approaches that of the isolated two-qubit subsystem as the spectator qubits become sufficiently detuned from the control and target qubits. Thus, we find that gate errors in multiqubit systems
can be reduced through sufficient detuning between fluxoniums, where frequency crowding is easier to avoid. 

We further evaluate gate performance as a function of gate time and find that the low-error regime (below $10^{-3}$) is achieved for gate times exceeding approximately 50 ns. Achieving such fast, high-fidelity gates requires sufficiently strong coupling between the fluxoniums, which in turn motivates a closer analysis of the parasitic ZZ interaction. This interaction introduces unwanted state-dependent energy shifts that degrade two-qubit gate fidelity. Ideally, the ZZ coupling strength should remain below $ 100$ kHz \cite{fors2024comprehensiveexplanationzzcoupling} to avoid significant degradation.

To quantify the impact of ZZ interactions in the presence of spectator qubits, we calculate the ZZ phase accumulation rate for the gate qubits with varying spectator-qubit states. We observe the maximum value of the ZZ interaction in the MLHM configuration, reaching approximately 25 kHz. 

We further extrapolate these results to longer chains of fluxonium qubits 
to identify frequency configurations that remain favorable as additional spectator qubits are introduced. Our analysis indicates that robust gate performance is achieved when spectator qubits are sufficiently detuned from both their immediate neighbors and the gate qubits, thereby minimizing unwanted multiqubit interactions arising from frequency crowding.

The paper is organized as follows. In Section~\ref{sec:II}, we define the system Hamiltonian, introduce the drive matrix elements, and discuss strategies for mitigating ZZ crosstalk. In Section~\ref{sec:III}, we analyze \textsc{CNOT} gates via selective darkening of transitions, and frequency configurations for both the two-qubit and four-qubit systems. Finally, Section~\ref{sec:IV} summarizes our results and offers perspectives on their implications for scalable fluxonium-based architectures.

\section{System model and its spectrum}
\label{sec:II}

\subsection{Circuit Hamiltonian}
\label{subsec:system}

We study a circuit consisting of four fluxonium qubits arranged in a linear chain with nearest-neighbor coupling mediated by kinetic mutual inductance, as shown in Fig.~\ref{fig:fluxsyst}.
The two central qubits implement a controlled-NOT (\textsc{cnot}) gate and are labeled as $F_2$ and $F_3$, while the two outer qubits act as spectators and are denoted $F_1$ and $F_4$.
This geometry reflects a minimal extension of a two-qubit device that allows us to quantify the impact of additional qubits on gate performance. We therefore describe the dynamics of the system using the Hamiltonian
\begin{equation}
\hat{H} = \sum_{\alpha} \hat{H}^{(0)}_\alpha + \hat{V} + \hat{H}_{\mathrm{drive}},
\label{eq:hamiltonian}
\end{equation}
where $\hat{H}^{(0)}_\alpha$ denotes the Hamiltonian of an individual fluxonium qubit with $\alpha= \{F_1,F_2,F_3,F_4\}$, $\hat{V}$ accounts for inter-qubit coupling, and $\hat{H}_{\mathrm{drive}}$ describes the local microwave drive. Each fluxonium qubit is modeled by the single-qubit Hamiltonian
\begin{equation}
\hat{H}^{(0)}_\alpha =
4E_{C_\alpha}\hat{n}_\alpha^2
+ \frac{1}{2}E_{L_\alpha}\hat{\phi}_\alpha^2
- E_{J_\alpha}\cos(\hat{\phi}_\alpha - \varphi_{\mathrm{ext}}),
\label{eq:fluxhamiltonian}
\end{equation}
where $E_{C_\alpha}$, $E_{L_\alpha}$, and $E_{J_\alpha}$ are the charging, inductive, and Josephson energies, respectively.
The operators $\hat{\phi}_\alpha$ and $\hat{n}_\alpha$ represent the superconducting flux and charge terms, which satisfy the canonical commutation relation $[\hat{\phi }_{\alpha}, \hat{n}_{\alpha}] = i$ \cite{Manucharyan_2009}.

Inter-qubit coupling is dominated by inductive interactions and is described by
\begin{equation}
\hat{V} =
J_{ff} \sum_{\langle \alpha,\beta \rangle} \hat{\phi}_\alpha \hat{\phi}_\beta
+ J_{nn} \sum_{\langle \alpha,\beta \rangle} \hat{n}_\alpha \hat{n}_\beta,
\label{eq:inthamiltonian}
\end{equation}
where the sums are restricted to nearest-neighbor qubit pairs.
In the device configuration considered here, capacitive coupling is suppressed throughout the analyses.
The consequences of a nonzero $J_{nn}$ for parasitic ZZ interactions are examined explicitly in Sec.~\ref{subsec:zzinteraction}. Finally, we introduce the time-dependent microwave drive, modeled by
\begin{equation}
\hat{H}_{\mathrm{drive}} =
\hbar \epsilon f(t)\cos(\omega_d t)
\left(\hat{n}_{F_2} + \eta \hat{n}_{F_3}\right),
\label{eq:drivehamiltonian}
\end{equation}
where $\epsilon$ sets the overall drive amplitude and $\eta$ controls the relative drive strength applied to the target qubit, which is chosen to satisfy the selective darkening condition.

The envelope function $f(t)$ implements a smooth pulse,
\begin{equation}
f(t)=
\begin{cases}
\sin^2\!\left(\dfrac{\pi t}{2t_r}\right), & t < t_r, \\
1, & t_r \le t < t_g - t_r, \\
\sin^2\!\left(\dfrac{\pi (t_g - t)}{2t_r}\right), & t \ge t_g - t_r ,
\end{cases}
\end{equation}
with total gate duration $t_g$ and rise time $t_r$.

Unless stated otherwise, all fluxonium qubits are taken to have identical charging and inductive energies,
$E_{C_\alpha}/h=1.0$~GHz and $E_{L_\alpha}/h=0.7$~GHz, and interact via an inductive coupling strength $J_{ff}/h=0.003$~GHz.
Qubit frequencies are differentiated solely through their Josephson energies, which are set to
$E_{J_\alpha}/h=4.5$, $3.8$, and $3.0$~GHz for the low (L), medium (M), and high (H) configurations, respectively. An extension of this notation is L', M', H', which denotes the original L-M-H value reduced by $0.1$ GHz. 
The impact of these frequency assignments on \textsc{cnot} gate performance is discussed in Sec.~\ref{subsec:energyandgateconfigs}.

The full system Hamiltonian is modeled and numerically diagonalized using the Python library, \textsc{scQubits}~\cite{Groszkowski2021scqubitspython, Chitta_2022}. For the numerical analysis we utilized at least five eigenstates of each qubit obtained from the time-independent Hamiltonian. These eigenstates are labeled $| \alpha k l \beta \rangle$, where $\alpha$ and $\beta$ denote the spectator qubits and k and l label the control and target qubits, respectively. The computational subspace is defined by $\alpha, \beta, k, l \in \{0,1\}$.

\subsection{ZZ Interaction}
\label{subsec:zzinteraction}
The parasitic ZZ interaction arises from static qubit–qubit coupling and induces unwanted conditional energy shifts, whereby the transition frequency of one qubit depends on the state of another. Such state-dependent shifts generate uncontrolled conditional phases during gate operations and constitute a significant limitation for achieving high-fidelity two-qubit gates in superconducting circuits \cite{xu2024modelingsuppressingunwantedparasitic, fors2024comprehensiveexplanationzzcoupling, dimitrov2025crossresonantgateshybridfluxoniumtransmon}.

In prior work on a two-fluxonium device, suppression of capacitive coupling enabled ZZ interactions as small as 2 kHz \cite{Lin_2025}. Extending this architecture to a chain of four fluxonium qubits introduces additional interactions, making the control of ZZ coupling an increasingly important challenge for scalability. Thus, to ensure that ZZ-induced phase accumulation does not dominate the gate error, we adopt a conservative threshold of $100$ kHz. 

In an extended fluxonium chain, the ZZ interaction is no longer characterized by a single value. Instead, it depends both on the specific qubit pair under consideration and on the states of the remaining qubits. Our analysis of ZZ therefore addresses two critical questions: first, whether inductive coupling remains effective at suppressing ZZ interactions in a multiqubit fluxonium chain; and second, to quantify the sensitivity of the ZZ interaction with the introduction of the spectator-qubit states.

We present the formula for the ZZ interaction between the control and target qubits while accounting explicitly for the neighboring spectator states $\alpha, \beta \in \{0,1\}$,
\begin{equation}
ZZ_{\alpha \beta}
= \frac{1}{h}\left(E_{|\alpha11 \beta\rangle} - E_{| \alpha 10 \beta\rangle} - E_{|\alpha 01 \beta\rangle} + E_{|\alpha 00 \beta\rangle}\right).
\end{equation}
The corresponding ZZ values for gate qubits and spectator states are summarized in Table ~\ref{tab:zz_interactions_configurations} for several representative frequency configurations, which are explained further in Sec. ~\ref{subsec:energyandgateconfigs}.

From these calculations, we find that the ZZ interaction is strongest for the LMHL' configuration and weakest for the HLMH' configuration. This behavior can be understood by recalling that the effective ZZ coupling originates from perturbative virtual transitions into noncomputational states. The energies of these intermediate states are set by the Josephson energies of the participating qubits, so different qubit pairings exhibit different levels of hybridization. As a result, configurations with more closely spaced energy levels produce stronger virtual mixing, whereas more widely separated levels suppress this effect.

\begin{figure}[htbp]
  \centering
  \includegraphics[width=0.5\textwidth]{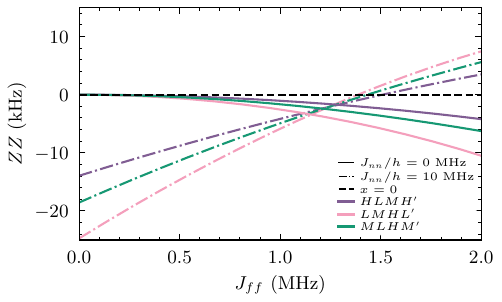}
  \caption{ZZ interaction strength for the HLMH', LMHL', and MLHM' frequency configurations, evaluated with spectator-qubit states $\alpha, \beta =1$, as a function of the inductive coupling constant $J_{ff}$. Results are shown both without capacitive coupling ($J_{nn} = 0.00$~GHz) and with a fixed capacitive contribution ($J_{nn} = -0.01$~GHz). The black horizontal line indicates the zero reference.}
  \label{fig:zz_barplot}
\end{figure}

To demonstrate the advantage of inductive coupling, we explicitly compare the inductive and capacitive contributions to this parasitic interaction in a four-qubit fluxonium chain for the HLMH', LMHL', and MLHM' configurations. For each configuration, we compute ZZ between the gate qubits while fixing the spectator-qubit states to $\alpha, \beta =1$. This choice is representative, as the ZZ interaction depends only weakly on the spectator states. Figure~\ref{fig:zz_barplot} presents the resulting ZZ interaction strengths as a function of the inductive coupling $J_{ff}$, shown both in the absence of capacitive coupling ($J_{nn}/h = 0$~MHz) and with a constant capacitive contribution ($J_{nn}/h = 10$~MHz).

When capacitive coupling vanishes, the ZZ interaction is minimized across the full range of inductive coupling strengths considered. Although fine-tuned combinations of inductive and capacitive coupling can, in principle, lead to cancellation of ZZ at isolated parameter values, maintaining such cancellation experimentally is hardly achievable. We therefore adopt a purely inductive coupling scheme, which showcases reduced ZZ interactions. Consistent with our earlier observations, the dependence of ZZ on the spectator-qubit states remains negligible in all cases considered.

\subsection{Matrix Elements}
\label{subsec:systemspectra}
We characterize the system’s response to external microwave fields by evaluating charge–matrix elements of the form $\langle \alpha 1 k \beta | \hat{n}_{Q} | \alpha 0 k \beta \rangle$, with $k = 0,1$. Within this notation, terms with $Q = F_2$ correspond to the direct drive of the control qubit, while those with $Q = F_3$ capture the cross-resonant drive acting on the target qubit. Analogously, matrix elements of the form $\langle \alpha k 1 \beta | \hat{n}_{Q} | \alpha k 0 \beta \rangle$ describe the direct drive for $Q = F_3$ and the cross drive for $Q = F_2$.

Across all frequency configurations considered, these matrix elements exhibit only weak dependence on the spectator-qubit states $\alpha$ and $\beta$. Their distinguishing feature is instead their relative sign: the direct-drive elements maintain a consistent sign, whereas the cross-resonant contributions appear with opposite parity. The corresponding values are summarized in Table~\ref{tab:zz_interactions_configurations}.

\begin{table*}
\centering
\begin{tabular}{|c|c|c|c|c|c|c|c|c|c|c|}
\hline
& \textbf{$\alpha \beta$}
& \textbf{$\mathrm{ZZ}_{\alpha\beta}$}
& \multicolumn{2}{c|}{$-i\bra{\alpha 10 \beta}\hat n_Q \ket{\alpha 00 \beta}$} & \multicolumn{2}{c|}{$-i\bra{\alpha 11 \beta}\hat n_Q \ket{\alpha 01 \beta}$}  
& \multicolumn{2}{c|}{$-i\bra{\alpha 01 \beta}\hat n_Q \ket{\alpha 00 \beta}$} & \multicolumn{2}{c|}{$-i\bra{\alpha 11 \beta}\hat n_Q \ket{\alpha 10 \beta}$} \\
\cline{6-7}
\cline{4-5}
\cline{8-9}
\cline{10-11}
&  & kHz & \textbf{$Q = F_2$} & \textbf{$Q = F_3$} & \textbf{$Q = F_2$} & \textbf{$Q = F_3$}
&\textbf{$Q = F_2$} & \textbf{$Q = F_3$} & \textbf{$Q = F_2$} & \textbf{$Q = F_3$}\\
\hline

\multirow{4}{*}{\textbf{HLMH'}}
 & 00 &  -9.5624 &  -0.08167 & 0.01217 & -0.08164 & -0.01188 & -0.01287 & -0.11603 & 0.01318 & -0.11607\\
 & 11 &  -9.5615 &  -0.08165 & 0.01215 & -0.08162 & -0.01189 & -0.01288 &  -0.116 & 0.01316 & -0.11604\\
 & 01 &  -9.5616 &  -0.08167 & 0.01215 & -0.08164 & -0.01189 & -0.01287 & -0.116 & 0.01318 & -0.11604\\
 & 10 &  -9.5623 &  -0.08165 & 0.01217 & -0.08162 & -0.01187 & -0.01288 &  -0.11603 & 0.01316 & -0.11607\\
\hline

\multirow{4}{*}{\textbf{LMHL'}}
 & 00 &  -24.7227 &  -0.11593 & 0.00852 & -0.1159 & -0.00803 & 0.00933& 0.16822 & -0.00969 & 0.16824  \\
 & 11 &  -24.7229 &  -0.11598 & 0.00862 & -0.11594 & -0.00796 & 0.00911 & 0.16822 & -0.00992 & 0.16829 \\
 & 01 &  -24.7121 &  -0.11591 & 0.00862 & -0.11589 & -0.00796 &0.00933 &0.1682 & -0.00969 & 0.16826 \\
 & 10 &  -24.7335 &  -0.11596 & 0.00852 & -0.11592 & -0.00804 & 0.00911 & 0.1682 & -0.00992 & 0.16822\\
\hline

\multirow{4}{*}{\textbf{MLHM'}}
 & 00 &  -12.9368 &  -0.08168 & 0.00442 & -0.08166 & -0.0039 & 0.00506 & 0.16817 & -0.00525 & 0.16817\\
 & 11 &  -12.9368 &  -0.08164 & 0.00429 & -0.08163 & -0.00401 & 0.00483 & 0.1682 & -0.00549 & 0.16824\\
 & 01 &  -12.9416 &  -0.08167 & 0.00429 & -0.08165 & -0.00401 & 0.00507 & 0.16819 & -0.00526 & 0.16822\\
 & 10 &  -12.9320 &  -0.08163 & 0.00442 & -0.08162 & -0.00389 & 0.00483 & 0.16816 & -0.00549 & 0.16816\\
\hline

\end{tabular}
\caption{ZZ interaction strengths and charge–matrix elements for the HLMH', LMHL', and MLHM' configurations, with control–target roles exchanged between $F_2$ and $F_3$. Values are reported for all spectator-qubit states $\alpha \beta$.}
\label{tab:zz_interactions_configurations}
\end{table*}


\section{\textsc{cnot} GATES}
\label{sec:III}
\subsection{\textsc{cnot} Gates via Selective Darkening of Transitions}
\label{subsec:cnotgates}
The controlled-NOT (\textsc{cnot}) gate acts on a pair of qubits, one designated as the control (C) and the other as the target (T).
 Its action can be represented by the unitary operator
\begin{equation}
\hat{U}_{\textsc{cnot}} = \ket{0_C}\bra{0_C} \otimes\hat1_T + \ket{1_C}\bra{1_C} \otimes \hat X_T,
\end{equation}
Where $\hat X_T$ is the Pauli-X operator acting on the target qubit. This gate flips the state of the target qubit when the control qubit is in the state $|1 \rangle$ and leaves it unchanged otherwise.

In this implementation, the \textsc{cnot} gate is realized via the cross-resonance (CR) technique, in which microwave pulses are applied directly to both the control and target qubits \cite{kang2025constructionnewtypecnot}. To realize the gate, the control and target qubits are driven at a frequency $\omega_d$ near the transition frequency of the target qubit. 
 
Both the control and target drives operate at a common resonance frequency, with amplitudes tuned to match the specific transition between qubit states. To this end, we employ a variation of the cross-resonance technique known as selective darkening (SD), which suppresses undesired transitions while enhancing the transition responsible for the \textsc{cnot} evolution \cite{de_Groot_2012}. Under the SD condition, unwanted transitions are rendered dark, meaning that the applied drive induces no population transfer between the corresponding states, while the desired transition remains resonantly driven.

 While this approach has been established for two fluxonium qubits~\cite{Nesterov_2022}, its extension to a four-qubit chain requires accounting for additional system eigenstates. We therefore label the static eigenstates of the whole system as $| \alpha k l \beta \rangle$, which reduce to bare product states in the limit of vanishing inter-qubit coupling. For fixed spectator states $\alpha, \beta$, the target transition is then confined to the subspace spanned by $\{\ket{\alpha k l\beta}\}$.

The selective-darkening (SD) gate suppresses the undesired transition 
$|\alpha 01 \beta \rangle \rightarrow |\alpha 0 0 \beta \rangle$ while keeping the $|\alpha 10 \beta \rangle \rightarrow |\alpha 11 \beta \rangle$ transition in the case of $F_2$ acting as the control and $F_3$ acting as the target:
\begin{subequations}
\begin{align}
\langle \alpha 0 1\beta| \hat{H}_{\text{drive}} |\alpha 0 0\beta \rangle &= 0, \\
\langle \alpha 1 0\beta| \hat{H}_{\text{drive}} |\alpha 1 1\beta \rangle &\neq 0.
\label{eq:nonzero}
\end{align}
\end{subequations}
Using $\langle \alpha 0 1\beta| \hat{H}_{\text{drive}} |\alpha 0 0\beta \rangle=0$, yields the condition for the ratio of drive amplitudes $\eta$:
\begin{equation}
\label{eq:eta}
\eta = -\frac{\langle \alpha 0 1\beta| \hat{n}_{F_2} |\alpha 00\beta \rangle}{\langle \alpha 0 1\beta| \hat{n}_{F_3} |\alpha 00\beta \rangle}.
\end{equation}

Assuming that the cross-drive matrix elements associated with $F_2$ and $F_3$ are equal in magnitude but opposite in sign, the effective matrix element governing the bright transition can be approximated as
\begin{equation}
2 (\eta \epsilon)  h|\langle \alpha 1 0\beta| \hat{n}_{F_3} |\alpha 1 1\beta \rangle|  f(t)\cos{(t)}.
\end{equation}
Under this approximation, larger values of $\eta$ provide the same transition matrix element for the desired transition at a weaker drive amplitude $\epsilon$.

We model the time evolution of the whole four-qubit device using QuTiP \cite{JOHANSSON20121760}. The simulation is performed in a truncated Hilbert space in which each fluxonium is restricted to a few ($N=5$) energy levels. After computing the full time evolution in this basis, we project the resulting evolution operator onto the computational subspace spanned by the $|\alpha k l \beta \rangle$ states.

To evaluate gate performance, we compared the projected evolution operator $\hat{U}$ with the ideal logical operation
\begin{equation}
\label{eq:ideal}
U_{\mathrm{id}}
= \hat{1}_{\mathrm{F}_1}
\otimes \textsc{CNOT}
\otimes \hat{1}_{\mathrm{F}_4}
\end{equation}

The gate error is then computed using the standard expression for a d-dimensional logical subspace
\begin{equation}
    \mathcal{E} = 1 - \frac{\mathrm{Tr}(\hat{U}^{\dagger}\hat{U})}{d(d+1)} - \frac{\left|\mathrm{Tr}(\hat{U}_{id}^{\dagger}\hat{U})\right|^2}{d(d+1)},
\end{equation}
where $d = 2^4$.

\subsection{Gate Optimization}
\label{subsec:energyandgateconfigs}

The frequency configuration is an important factor governing the performance of our four-fluxonium architecture. We adopt an L–M–H (Low, Medium, High) labeling convention to classify the three possible qubit frequencies that a fluxonium device may take within our two- or four-qubit chain. This convention allows us to examine how frequency detunings between neighboring qubits, particularly between the gate qubits and their spectators, influence gate performance. The labels correspond to the $|0\rangle$ and $|1 \rangle$ transition, with the H label indicating the largest transition frequency. Since this frequency is inversely related to the Josephson energy $E_J$, this parameter effectively characterizes the energy configuration of a fluxonium for our analysis, as all other circuit parameters are held fixed (see Sec.~\ref{sec:II}). To elucidate how these arrangements affect gate errors and to identify regimes that consistently yield error rates below $10^{-3}$, we first examine the behavior of a simpler two-fluxonium subsystem.

A two-fluxonium chain contains three distinct frequency configurations: LM, MH, and LH. In this notation, the control qubit $F_2$ and the target qubit $F_3$ are each assigned a frequency label from the set $\{L, M, H\}$, with the constraint that the two qubits do not share the same frequency in a given configuration. Each configuration is evaluated in two ways: once using $F_2$ as the control and $F_3$ as the target, and once with these roles reversed. 

A similar procedure is applied to the four-fluxonium chain. The two central qubits, which form the active gate pair, are assigned one of the LM, MH, or LH frequency configurations. The spectator qubits on either side are placed at a distinct frequency from the gate pair; both spectators are set to the same nominal value, with the right spectator detuned from the left by 0.1 GHz. A system with a given frequency configuration was also evaluated twice to account for the reversal of the control and target roles.
Table~\ref{tab:optimizedparams} provides a summary of the gate performance obtained across all frequency configurations and control–target assignments for both the two-qubit and four-qubit systems.
\begin{table}

\centering
\begin{tabular}{|c|c|c|c|c|}
\hline
\textbf{Configuration} & \textbf{$\epsilon$/$\hbar$ [GHz]} & \textbf{$\eta$} & \textbf{$\Delta$} & \textbf{Error} \\
\hline
\rowcolor{gray!15}
L\underline{\textbf{M}} & 0.396 & -0.107 & 1.000 & $2.59 \times 10^{-7}$ \\
HL\underline{\textbf{M}}H' & 0.397 & -0.107 & 1.000 & $4.45 \times 10^{-6}$ \\
\rowcolor{gray!15}
\underline{\textbf{L}}M & 0.423 & 0.431 & 1.000 & $4.92 \times 10^{-7}$ \\
H\underline{\textbf{L}}MH' & 0.425 & 0.146 & 0.998 & $4.05 \times 10^{-6}$ \\
\rowcolor{gray!15}
M\underline{\textbf{H}} & 0.543 & -0.272 & 1.000 & $7.71 \times 10^{-8}$ \\
LM\underline{\textbf{H}}L' & 0.552 & -0.267 & 1.000 & $7.84 \times 10^{-4}$ \\
\rowcolor{gray!15}
\underline{\textbf{M}}H & 0.620 & 0.205 & 1.000 & $1.20 \times 10^{-7}$ \\
L\underline{\textbf{M}}HL' & 0.623 & 0.0717 & 0.999 & $3.65 \times 10^{-5}$ \\
\rowcolor{gray!15}
L\underline{\textbf{H}} & 1.008 & -0.0281 & 0.999 & $8.62 \times 10^{-8}$ \\
ML\underline{\textbf{H}}M' & 1.034 & -0.142 & 0.999 & $5.10 \times 10^{-4}$ \\
\rowcolor{gray!15}
\underline{\textbf{L}}H & 1.205 & 0.0526 & 0.999 & $1.15 \times 10^{-7}$ \\
M\underline{\textbf{L}}HM' & 1.236 & 0.0520 & 0.999 & $1.16 \times 10^{-4}$ \\
\hline
\end{tabular}
\caption{Optimized gate parameters $\epsilon$, $\eta$, and $\Delta$, and the resulting gate error for the two-fluxonium chain under the LM, MH, and LH frequency configurations. Each configuration is evaluated for both control–target orientations, with the target qubit indicated in bold and underlined. The corresponding four-fluxonium results are listed immediately below each two-fluxonium entry.}
\label{tab:optimizedparams}
\end{table}

We find that the re-optimized gate parameters vary minimally between the two-fluxonium system and its four-fluxonium counterpart. This variation indicates that when the spectator qubits are sufficiently detuned from the gate qubits, the performance of the four-fluxonium chain closely resembles that of the corresponding two-qubit subsystem. As a consequence, the optimized parameters obtained from the no-spectator model provide a reliable estimate for the gate error in the extended chain. Even without re-optimization, these parameters yield reasonable error predictions, while re-optimization of the multiqubit system further reduces the error below the $10^{-3}$ threshold.

\begin{figure}[htbp]
  \centering
  \includegraphics[width=0.5\textwidth]{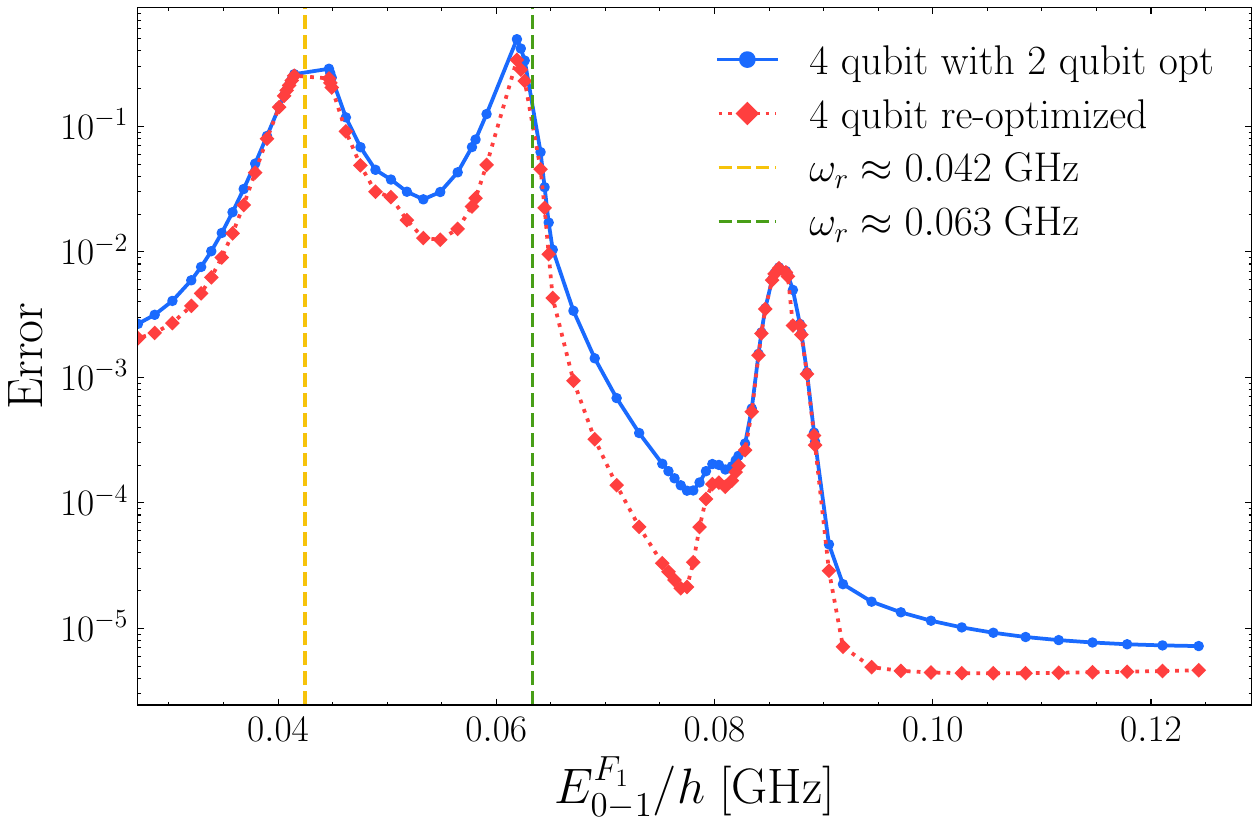}
  \caption{Error threshold for a four-qubit system with the SLMS' configuration, obtained by sweeping the spectator-qubit frequency $E_{0-1}^{F_1}/h$. The unoptimized gate errors, shown in blue, are computed using the optimized two-fluxonium gate parameters from the LM configuration. The optimized gate errors are shown in red. The first resonance condition, where the spectator qubits approach the L frequency, is indicated by the yellow line. The second resonance condition, where the spectator qubits approach the M frequency, is indicated by the green line.}
  \label{fig:optvunopt}
\end{figure}

Figure \ref{fig:optvunopt} illustrates both the effect of detuning relative to the gate qubits and the comparison between optimized and unoptimized gate parameters in a four-fluxonium chain. The example shown corresponds to the SLMS' frequency configuration, in which the two central qubits serve as the control and target (in that order), and the outer qubits serve as spectators. In this analysis, the spectator frequency is swept over a broad range of values while the resulting gate error is computed for each setting. The unoptimized four-qubit errors (blue curve) are obtained using the parameters optimized for the two-fluxonium LM configuration, whereas the fully optimized four-fluxonium errors are shown in red. Vertical lines mark the resonance conditions at which a spectator frequency coincides with that of either the control or the target qubit.
The gate error is maximized when the spectator qubits are near resonance with the gate qubits, exceeding an error threshold of $10^{-1}$. As the spectators are detuned away from the gate qubits, the error is strongly suppressed, falling below $10^{-5}$.

\begin{figure}[htbp]
  \centering
  \includegraphics[width=0.5\textwidth]{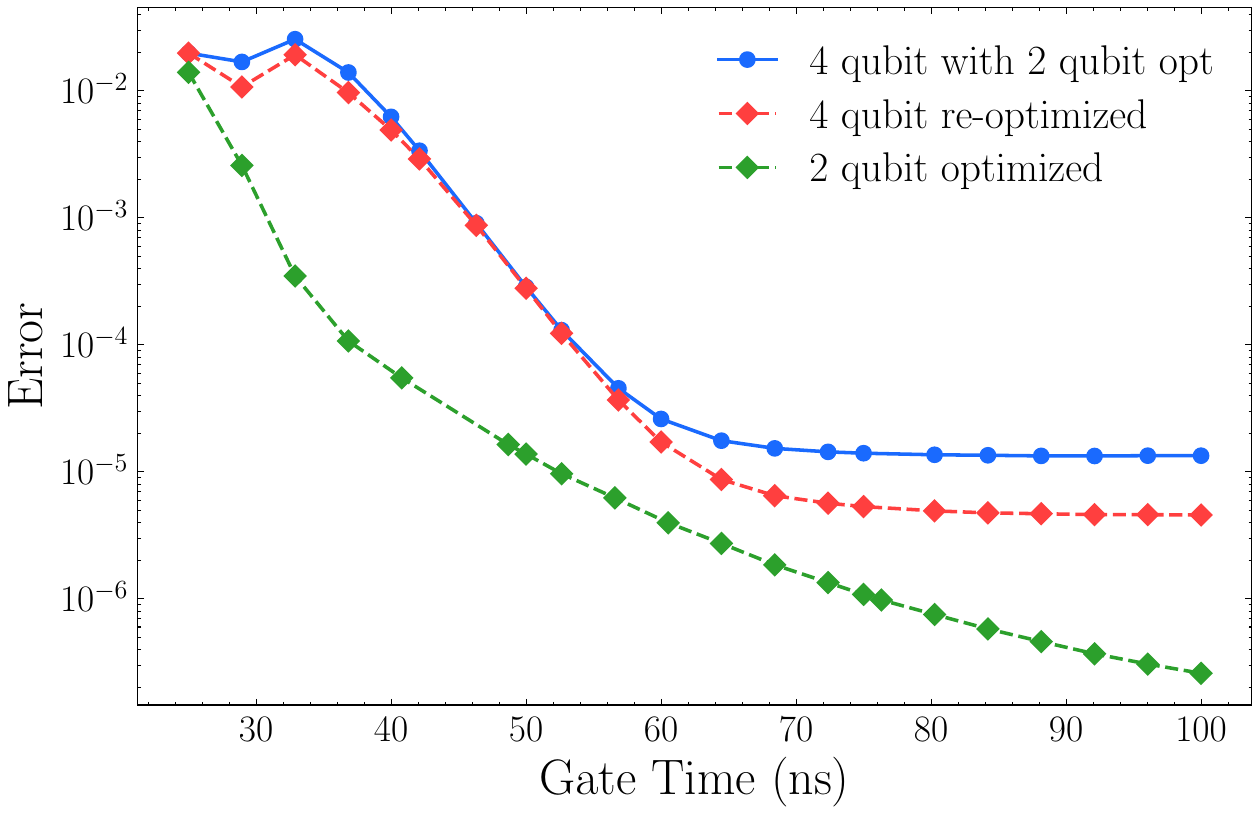}
  \caption{Gate error with respect to gate time for an SLMS' frequency configuration with the spectator Josephson energy valued at 3.05 GHz.}
  \label{fig:3.5gatetime}
\end{figure}

The gate performance can also be assessed as a function of the gate duration, as shown in Fig.~\ref{fig:3.5gatetime}. By choosing a configuration in which the spectator qubit $F_1$ has a Josephson energy of $E_{J}/h= 3.05$~GHz we find that high-fidelity operations are preserved even at shorter gate times. We notice this low error trend for both the optimized four-qubit drive and the unoptimized four-qubit drive using the original two-qubit LM configuration drive parameters.

\section{Discussion and Conclusions}
\label{sec:IV}
In this work, we investigated how frequency configurations affect the error rates of a \textsc{cnot} gate within a chain of inductively coupled, nearest-neighbor fluxonium qubits. We established a baseline using two-qubit subsystems in low-medium, medium-high, and low-high configurations, achieving gate errors below $10^{-5}$ for a gate time of 100 ns. These results provided a reference for understanding how additional qubits reduce gate performance. By introducing spectator qubits adjacent to the control and target qubits, we quantified the increase in gate error rate caused by those spectators. Still, we report error values below $10^{-4}$ for 100 ns gates and below $10^{-3}$ for 50 ns gates, demonstrating that high-fidelity operations are achievable in multi-fluxonium systems. Crucially, when spectator qubits are far detuned from the active gate qubit pair, the four-fluxonium system reaches the behavior of the isolated two-fluxonium system. However, as spectator frequencies approach the resonant conditions of the gate qubits, hybridization leads to a rapid deterioration of the gates. Furthermore, we demonstrate that the optimized pulse parameters derived from gate design for an isolated qubit pair can serve as a diagnostic tool before committing to complete optimization of the multiqubit system. If the gate fidelity optimized for the pair drops significantly in the presence of spectators, whole-system optimization is unlikely to yield high performance. Conversely, if the optimized two-fluxonium configuration is retained, whole system optimization can yield moderate improvement. 

Finally, because the coupling between neighboring fluxoniums is short-range, these principles naturally extend to larger fluxonium systems. By repeating the quasi-periodic frequency pattern shown in Fig.~\ref{fig:fluxsyst}, with adjacent qubits appropriately detuned, one can construct a multiqubit architecture in which high-fidelity \textsc{cnot} gates can be implemented between each neighboring pair. Such a design provides a promising pathway toward scalable fluxonium-based quantum processors with low gate error rates and reduced ZZ.

\begin{acknowledgments}
We are thankful to Jiakai Wang for insightful discussions. Some parts of our numerical simulations were performed using the QuTiP~\cite{JOHANSSON20121760, JOHANSSON20131234} and scQubits~\cite{Groszkowski2021scqubitspython, Chitta_2022} python packages.
\end{acknowledgments}

\bibliography{main}

\end{document}